\documentclass[doublecol]{epl2} 

\usepackage{textcomp}
\usepackage{makeidx}
\usepackage{amsmath}
\usepackage{subfigure}
\usepackage{amssymb}
\usepackage{hyperref}
\usepackage{graphicx}
 \usepackage{epstopdf}
\usepackage[utf8]{inputenc}
\usepackage{float}
\usepackage{color}
\usepackage{epstopdf}
\title{A new model of regular black hole in $(2+1)$ dimensions}
\shorttitle{A New RBH in $(2+1)$} 

\author{Milko Estrada \inst{1} \and Francisco Tello-Ortiz \inst{1}}
\shortauthor{M.Estrada \etal}

\institute{  
  \inst{1}Departamento de F\'isica, Facultad de Ciencias Básicas, Universidad de Antofagasta, Casilla 170, Antofagasta, Chile.
}

\abstract{
We provide a new regular black hole solution in $(2+1)$ dimensions with presence of matter fields in the energy momentum tensor. The inclusion of our proposed energy density  plus a negative cosmological constant allows that the solution can have both flat as de--Sitter or Anti de--Sitter core. This latter is a proper characteristic of our solution, because other models of regular black holes have only a single type of core.
 Since the first law of thermodynamics for regular black holes is modified by the presence of the matter fields, we provide a new version of the first law, where a local definition of the variation of energy is defined, and, where the entropy and temperature are consistent with the previously known in literature. At the hypothetical limit when the horizon radius $r_+ \to \infty$ the usual first law $dM=TdS$ is recovered. The effectiveness of the formalism used to compute the mass of our regular black holes in $(2+1)$ suggests the potential applicability of this method to calculate the mass of other models of regular black holes in $d \ge 4$ dimensions.}

\begin{document}

\maketitle

\section{Introduction}

The recent detection of gravitational waves through the collision of two rotating black holes \cite{Abbott:2016blz,Abbott:2017oio}, together with the also recent assignment of the Nobel Prize, 
 have positioned the black holes as one of the most interesting and intriguing objects in gravitation. In this connection, the fact that the black holes, due to quantum fluctuations emit as black bodies, where its temperature is related to its surface gravity \cite{Hawking:1974sw}, shows that in these objects the geometry and thermodynamics are directly connected. 

The Schwarzschild black hole has a central singularity where the laws of physics cease to operate. From the classical point of view, Bardeen in reference \cite{bardeen} proposed the first model of regular black holes (RBHs). In this model the singularity is avoided by changing the mass parameter in the Schwarzschild solution by a radial mass function such that near the origin the function behaves as $f \approx 1-Cr^2$, {\it i.e.} the solution has a de sitter core. After this,  several models of regular black holes have been constructed, where the Einstein field equations are coupled to non--linear electromagnetic sources. Examples of this can lead to RBHs with de--Sitter core \cite{AyonBeato:1998ub} or flat core \cite{Culetu:2014lca,Balart:2014cga}. 

Furthermore, it is possible to construct RBHs solutions including matter fields in the energy momentum tensor. Examples of this are the models of references \cite{Dymnikova:1992ux,Hayward:2005gi}. These models are solutions of General Relativity and have de--Sitter core. We can see a review of these types of RBHs in reference \cite{Spallucci:2017aod}. Other models of RBHs based in the presence of matter fields in the energy momentum tensor for higher curvature theories, with a de--Siter core, can be found in references \cite{Aros:2019auf,Aros:2019quj,Babichev:2020qpr}. 

It is well known that the first law of thermodynamics is modified for RBHs due to the presence of matter fields in the energy momentum tensor. In reference \cite{Ma:2014qma} it have been shown a way of writing the first law including a correction factor, which corresponds to an integration of  the radial coordinate up to infinity. However, the first law of thermodynamics for RBHs is still an open question of physical interest.

On the other hand, models of gravity in $(2+1)$ dimensions have drawn high attention in the last years, due to the simplicity of its equations of motion. These models can be viewed as toy models, with the hope that its results can help the understanding of $(3+1)$ and higher dimensional models of gravity \cite{Astefanesei:2003rw}. Models of black holes in $(2+1)$ also have been widely studied in the last years. BTZ model \cite{Banados:1992wn,Banados:1992gq} is a vacuum black hole solution of the Einstein field equations in $(2+1)$ dimensions, which has been studied to find some conceptual issues in quantum gravity, string theory and AdS/CFT correspondence \cite{Hendi:2015bba}. In this solution the curvature invariants are regular everywhere. Regular black holes in $(2+1)$ dimensions coupled to a non--linear electrodynamics were studied in reference \cite{Cataldo:2000ns,He:2017ujy}. Nevertheless, in the literature also there are solutions in $(2+1)$ dimensions with a central singularity, as for example the solutions of reference \cite{Oliva:2009ip} for Massive Gravity \cite{Bergshoeff:2009hq}, or the solution given in \cite{Sa:1996ty}. See references \cite{Hendi:2012zz,Hendi:2017oka,Hendi:2017mgb}  for a generalization of BTZ model with other fields and references \cite{Hendi:2010px,Hendi:2020yah} for an interesting higher dimensional generalization of BTZ and geodesic analysis. We can see more recent studies in references \cite{Casals:2019jfo,Tang:2019jkn,Kaur:2018tzl,Bussola:2018iqj,Zeng:2019huf,Sheykhi:2020dkm}.

Thus, it is undoubtedly interesting the seeking of new black hole solutions in $(2+1)$ dimensions with matter fields in the energy momentum tensor and the study of its thermodynamics properties. In this work we will provide a new model of energy density in $(2+1)$ dimensions, which will lead to a new regular black hole solution. Furthermore we will propose a structure for the first law of thermodynamics for RBHs and compute the mass of the solution using a definition of conserved charged appeared recently in the literature \cite{Aoki:2020prb,Sorge:2020pdj}. Furthermore we will study the stability of our solution.

\section{$(2+1)$ Einstein Field Equations}

The Einstein field equation are given by:
\begin{equation} \label{EcuacionDeEinstein}
    G^\mu_{\,\,\,\nu}+\Lambda \delta^\mu_{\,\,\,\nu}= 8 \pi \bar{G} T^\mu_{\,\,\,\nu},
\end{equation}
where the cosmological constant is equal to $\Lambda = -\dfrac{1}{l^2}$. The Newton constant, in the natural system of units, has units of length $[\bar{G}]=\ell$. For simplicity, we consider arbitrarily that this constant has a magnitude equal to 1. The energy momentum tensor describing a perfect fluid, is given by:
\begin{equation}
    T^\mu_{\,\,\,\nu}=\mbox{diag}(-\rho,p_r,p_\theta).
\end{equation}

We shall study the following spherically symmetric space time:
\begin{equation} \label{elementodelinea1}
    ds^2=-f(r)dt^2+f(r)^{-1}dr^2+r^2 d\theta. 
\end{equation}

 This form of the metric has the following consequence on the energy momentum tensor through the Einstein field equations:
\begin{equation}
    \rho=-p_r.
\end{equation}

So, for the line element \eqref{elementodelinea1}, the $(t,t)$ and $(r,r)$ component of the Einstein field equations are:
\begin{equation} \label{ComponentesEinstein}
    \frac{1}{2r} \frac{df}{dr} + \Lambda =-8 \pi \rho
\end{equation}

On the other hand, from the conservation law, $\nabla_\mu T^{\mu \nu}=0$, one gets
\begin{equation}
    p_\theta = r \frac{d}{dr}p_r+p_r.
\end{equation}

We will chose the following ansatz:
\begin{equation} \label{elementodelinea2}
    f(r)=1-8\bar{G}m(r)-\Lambda r^2.
\end{equation}

Next, replacing the ansatz \eqref{elementodelinea2} into the equation \eqref{ComponentesEinstein}, one arrives to:
\begin{equation}
 \frac{d}{dr} m(r)= 2 \pi r \rho(r).
\end{equation}

\section{The new model}

The mass function for the $(2+1)$ case is defined as:
\begin{equation} \label{funciondemasa}
    m(r)=2\pi \int_0^r x \rho (x) dx .
\end{equation}

As was mentioned above, in this work we will provide a new RBH solution based in the inclusion of matter field in the energy momentum tensor. For this, we will propose a new model of energy density. In order to have a well posed physical situation, the energy density model that we will choose, follows the requirements described in reference \cite{Aros:2019quj}:
\begin{itemize}
    \item The energy density must be positive and continuously differentiable to avoid singularities.
    \item The energy density must have a finite single maximum at the origin. The finiteness of $\rho(0)$ forbids the presence of a central singularity.
    \item In order to guarantee a well defined asymptotic behavior, the energy density must be a decreasing radial function and must vanish at infinity. Furthermore, the mass function must reach its finite maximum value at infinity, {\it i.e}
    \begin{align}
        \displaystyle \lim_{r \to \infty} \rho &=0 \label{limiterho} \\
        \displaystyle \lim_{r \to \infty} m(r) &= \mbox{constant}.
    \end{align}
\end{itemize}

Thus, we propose the following model for energy density:
\begin{equation} \label{DensidaddeEnergia11}
    \rho = \frac{2LM^2}{\pi (2LM+r^2)^2},
\end{equation}
and replacing into the equation \eqref{funciondemasa} we obtain the mass function
\begin{equation} \label{funciondemasa1}
    m(r)= \frac{Mr^2}{2LM+r^2},
\end{equation}
where $L$ is a positive constant of units $[L]=\ell^3$. This constant must be positive to ensure that $\rho(0)=1/(2\pi L)>0$. The parameter $M$, is the so called mass parameter, whit units of $[M]=\ell^{-1}$. 
This constant also must be positive to ensure the absence of singularities in the energy density. 

Considering the previous analysis, the mass function has units $[m(r)]=\ell^{-1}$ and since $[\bar{G}]=\ell$, the factor $\bar{G}m(r)$ is dimensionless, where the magnitude of $\bar{G}$ is equal to the unity.

This model could be viewed as an extension in $(2+1)$ dimensions of the higher dimensional model of reference \cite{Aros:2019quj}, which for the $(3+1)$ dimensional case, coincides with the Hayward metric \cite{Hayward:2005gi}.

For the Hayward metric in $(3+1)$ dimensions, when the energy density is of the order of the Planck units near the origin, the formation of a de--Sitter core is associated with quantum fluctuations. These models are known as Planck Stars \cite{DeLorenzo:2014pta,Rovelli:2014cta}. In reference \cite{Rovelli:2017zoa}, using radio astronomy data, it is conjectured that Planck Stars represent a speculative but realistic possibility to testing quantum gravity effects.
Thus, if our energy density were of order of the Planck units near the origin, our model could
serve as a toy model to study these ideas in a future work
due to the simplicity of the equations of motion in $(2+1)$ dimensions. As we shall see below, the inclusion of a negative cosmological constant not only provides the formation of a dS core, but also provides the formation of a flat/AdS core. This latter feature could be analyzed for Planck Stars models elsewhere. It is worth mentioning that the formation of a flat or AdS core is a new feature for RBHs without non--linear electromagnetic sources.

The solution is: 
\begin{equation} \label{solucion}
    f(r)=1-\Lambda r^2 - 8 \bar{G} m(r)= 1 + \frac{r^2}{l^2} - \frac{8Mr^2}{2LM+r^2} .
\end{equation}

For this solution, the Ricci and Kretschmann invariants are given by:
\begin{equation}
    R=-\frac{6}{l^2}+\frac{48M}{2LM+r^2}-\frac{112Mr^2}{(2LM+r^2)^2}+\frac{64Mr^4}{(2LM+r^2)^3} ,
\end{equation}

\begin{align}
    K &= \left ( \frac{2}{l^2}-\frac{16M}{2LM+r^2} +  \frac{80Mr^2}{(2LM+r^2)^2} - \frac{64Mr^4}{(2LM+r^2)^3} \right )^2 \nonumber \\
    &+ 2 \left ( \frac{2}{l^2} - \frac{16M}{2LM+r^2} + \frac{16Mr^2}{(2LM+r^2)^2} \right)^2 .
\end{align}

We understand by regular when the curvature invariants are free of singularities. In our case both Ricci and Kretschmann don't diverge at $r=0$ nor other value of $r$ due that $L>0$ and $M>0$.

\subsection{behavior near the origin}

From equation (\ref{solucion}) it is evident that this function behaves near the origin as:
\begin{equation}
    f \big |_{r \approx 0} \approx 1 + \left ( \frac{1}{l^2} - \frac{4}{L} \right) r^2,
\end{equation}
thus, we can define the effective radius ($\tilde{l}$):
\begin{equation} \label{radioefectivo}
\frac{1}{\tilde{l}}= \left ( \frac{1}{l^2} - \frac{4}{L} \right)      
\end{equation}
\begin{itemize}
    \item For $\frac{1}{l^2} =\frac{4}{L}$ the behavior near the origin corresponds to a flat space time. Regular solutions with a flat core have been studied for $(3+1)$ dimensions with a non--linear electromagnetic source in reference \cite{Balart:2014cga,Culetu:2014lca}. However, for the $(2+1)$ case, with a nonzero energy density, this behavior is a proper feature of our solution.
    \item For $\frac{1}{l^2} <\frac{4}{L}$ the behavior near the origin corresponds to a dS space time. Several models of regular black holes share this feature \cite{Hayward:2005gi,Aros:2019auf,Aros:2019quj,DeLorenzo:2014pta,Rovelli:2014cta,Babichev:2020qpr}.
    In this case the equation \eqref{radioefectivo} represents an effective dS radius.
    
    \item For $\frac{1}{l^2} >\frac{4}{L}$ the behavior near the origin corresponds to an AdS space time. So, in our model this possibility could be valid, which differs from the most of the RBHs models. Due to the simplicity of the $(2+1)$ models, the physical consequences of this particular feature could be testing elsewhere.
   In this case the equation \eqref{radioefectivo} represents an effective AdS radius.
\end{itemize}

Thus, our model of energy density added to a negative cosmological constant allow that the solution can have both flat as de--Sitter or Anti de--Sitter core, depending on the values of the $l$ and $L$. This latter is a proper characteristic of our solution, because other models of regular black holes have only a single type of core.

\section{Thermodynamic analysis}

\subsection{Conserved Charges}

Recently, in references \cite{Aoki:2020prb,Sorge:2020pdj}  has appeared a new definition of conserved charges. In \cite{Aoki:2020prb} the energy and momentum can be computed by integrating a covariantly conserved current $J^\mu=T^\mu_\nu \xi^\nu$ in a volume integral. It is worth to mention that the definition of reference \cite{Aoki:2020prb} is reduced to the definition of conserved energy of reference \cite{Sorge:2020pdj} for a Killing vector $\xi^\mu=-\delta^\mu_0$.  Following \cite{Aoki:2020prb}, the energy is defined as:
\begin{equation} \label{CargaConservada}
    E = \int d^{d-1}x \sqrt{-g} J^0 = \int d^{d-1}x \sqrt{-g} T^0_\nu \xi^\nu,
\end{equation}
where $\xi^\nu$ is a Killing vector and $d$ is the number of dimensions. 

It is worth mentioning that this definition is diffeo-invariant. Furthermore the current is covariantly conserved due that $\nabla_\mu T^\mu_\nu=0$ and $\nabla_\mu \xi_\nu+\nabla_\nu \xi_\mu=0$.

Following \cite{Aoki:2020prb}, we choose $\xi^\mu=-\delta^\mu_0$. Evaluating our solution into the equation \eqref{CargaConservada}:
\begin{equation}
 \displaystyle   E=-2\pi \int^\infty_0 r dr T^0_0= 2\pi \int^\infty_0 r dr \rho(r)= \lim_{r \to \infty} m(r)=M
\end{equation}
Thus, the parameter $M$ represents the total energy of the black hole. On the other hand, the obtained mass $M$ coincides with the value of the vacuum BTZ solution computed in reference \cite{Miskovic:2006tm} using other definition of conserved charges.

So, it is computed the mass of the black hole in $(2+1)$ without adding boundary terms as in references \cite{Miskovic:2006tm,Miskovic:2016mvs}. The effectiveness of this formalism for computing the mass of our $(2+1)$ RBHs model could make us think about using this formalism for computing the mass of another models of regular black holes in $d \ge 4$ dimensions, where the energy density also fulfills the conditions mentioned before. Some examples of these models are \cite{Hayward:2005gi,DeLorenzo:2014pta,Rovelli:2014cta,Dymnikova:1992ux,Aros:2019quj}.

\subsection{The first law of thermodynamic in $(2+1)$ dimensions with matter fields}

It is well known that the first law of thermodynamics is modified for RBHs due to the presence of matter fields in the energy momentum tensor. Following \cite{Liang:2017kht,Balart:2019uok} we define the thermodynamics pressure as $p=- \dfrac{\Lambda}{8\pi}$.
In order to propose a structure for the first law of thermodynamics for RBHs, we will use the conditions $f(r_+,M,p)=0$ and $\delta f(r_+,M,p)=0$ , which can be viewed as constraints on the evolution along the space parameters \cite{Aros:2019auf,Zeng:2019huf}. Thus, from the condition
\begin{equation}
    0 = \frac{\partial f}{\partial r_+} dr_+ + \frac{\partial f}{\partial M} dM + \frac{\partial f}{\partial p} dp 
\end{equation}
it is straightforward to check that for the present solution \eqref{solucion}, the first law takes the form:
\begin{equation}
    \frac{\partial m}{\partial M} dM = \left ( \frac{1}{4\pi} f'\big |_{r=r_+}  \right ) d \left ( \frac{\pi r_+}{2} \right ) + (\pi r_+^2) dp.
\end{equation}

The above equation can be rewritten as:
\begin{equation} \label{duTdS}
du=TdS+ V dp  ,
\end{equation}
where the temperature and entropy are easily computed as: 

\begin{align} 
    T&= \frac{1}{4\pi} f'\big |_{r=r_+} \\
    S&= \frac{\pi r_+}{2} \\
    V&= \pi r_+^2,
\end{align}
where our definition of volume coincide with \cite{Liang:2017kht}.

One can notice that $p$, the pressure, has units of $[p]=\ell^{-2}$, equation \eqref{solucion}. Likewise, one can check that the factor $[\bar{G}m] = \ell^0$ is dimensionless and $[S] = \ell^{1}$.  Following {\it Euler's theorem} \cite{Altamirano:2014tva}, one can construct the {\it Smarr formula}:

\begin{equation}
    TS-2pV=0
\end{equation}
which coincide with the vacuum $(2+1)$ solution of reference \cite{Liang:2017kht} for the non rotating case. In a future work, could be studied the case where the constant $L$ is a thermodynamics parameter.

It is direct to check that the temperature for our model \eqref{solucion} is:
\begin{equation}
T= \frac{2}{r_+}m(r_+) - \frac{2}{r_+} - 8 \frac{dm}{dr}\Big |_{r=r_+} .
\end{equation}

On the other hand, the heat capacity at constant pressure is computed as:
\begin{equation}
    C_p=T \frac{dS}{dT} \bigg |_{p=cte} =T \frac{dS}{dr_+} \left( \frac{dT}{dr_+}  \right )^{-1} \bigg |_{p=cte}
\end{equation}

In the vacuum case, the first law is of the form $dM=TdS$, where $M$ is the energy of the system. However in our case the left side of the first law is modified in order to obtain the correct value of the entropy due to the presence of matter fields in the energy momentum tensor. So, in our case, the term of the left side, $du$, corresponds to a local definition of the variation of the energy. 

The modification of the first law for regular black holes was studied in reference \cite{Ma:2014qma}, without our constraint in the space of parameters, by including a correction factor which corresponds to an integration of the radial coordinate up to infinity. Unlike this latter reference, in our case both terms, $du$ and $dS$, are local variables defined at the horizon. 

It is easy to check that the factor $dm/dM$ in equation \eqref{duTdS} is always positive, and thus, the sign of the variation of $du$ always coincides with the sign of the variation of the total energy $dM$. Furthermore, 
at infinity it is fulfilled that $\displaystyle \lim_{r \to \infty} dm/dM=1$, therefore the variation of the local and total energy are similar at the asymptotically region. Thus, at the hypothetical limit when the horizon radius $r_+ \to \infty$, the usual first law $dM=TdS$ is recovered.

In a future work, following \cite{Hendi:2015wxa}, we could test a possible relation between $l$ and $L$ to study the ensemble dependency problem. 

\subsection{Stability of our solutions}

\begin{figure} 
   \begin{center}
      \includegraphics[width=0.32 \textwidth]{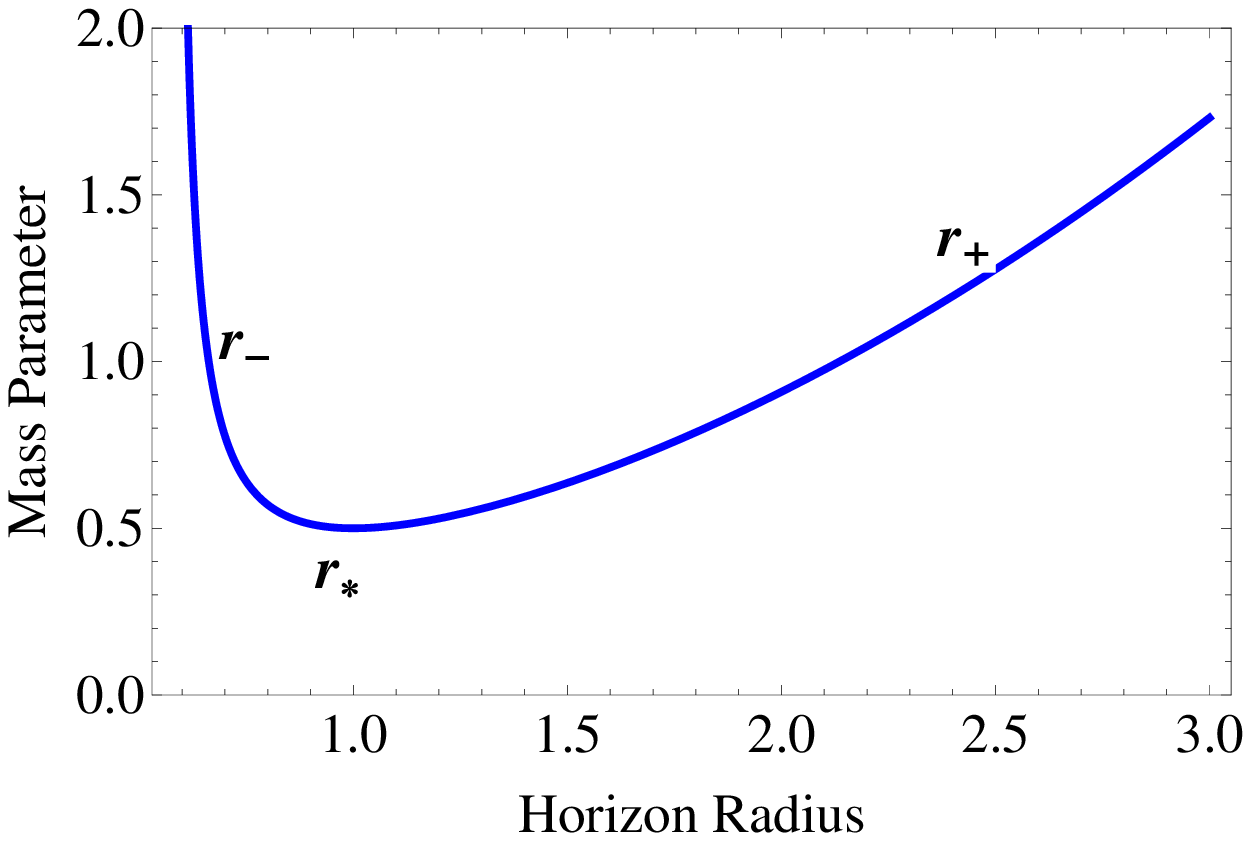} \ 
      \includegraphics[width=0.32 \textwidth]{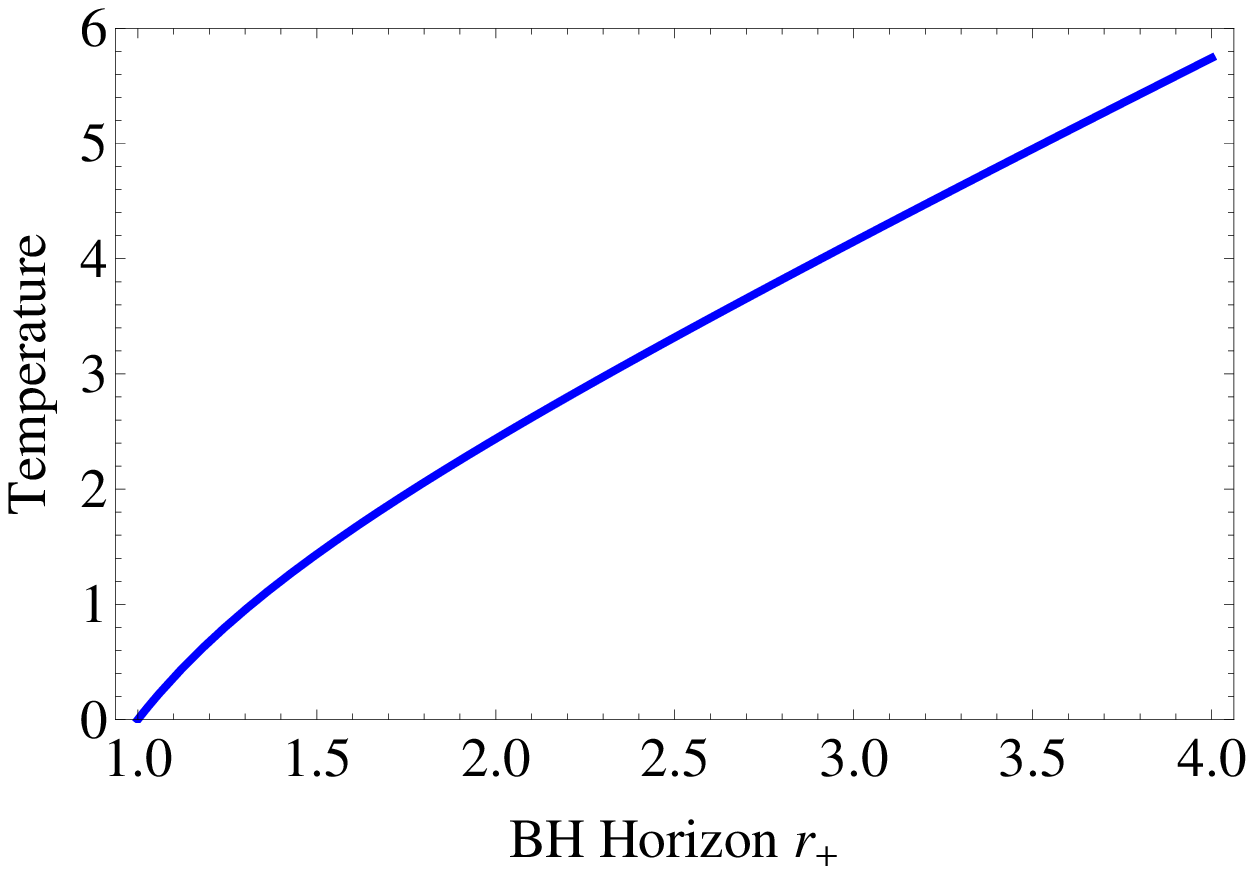} \ 
      \includegraphics[ width=0.32 \textwidth]{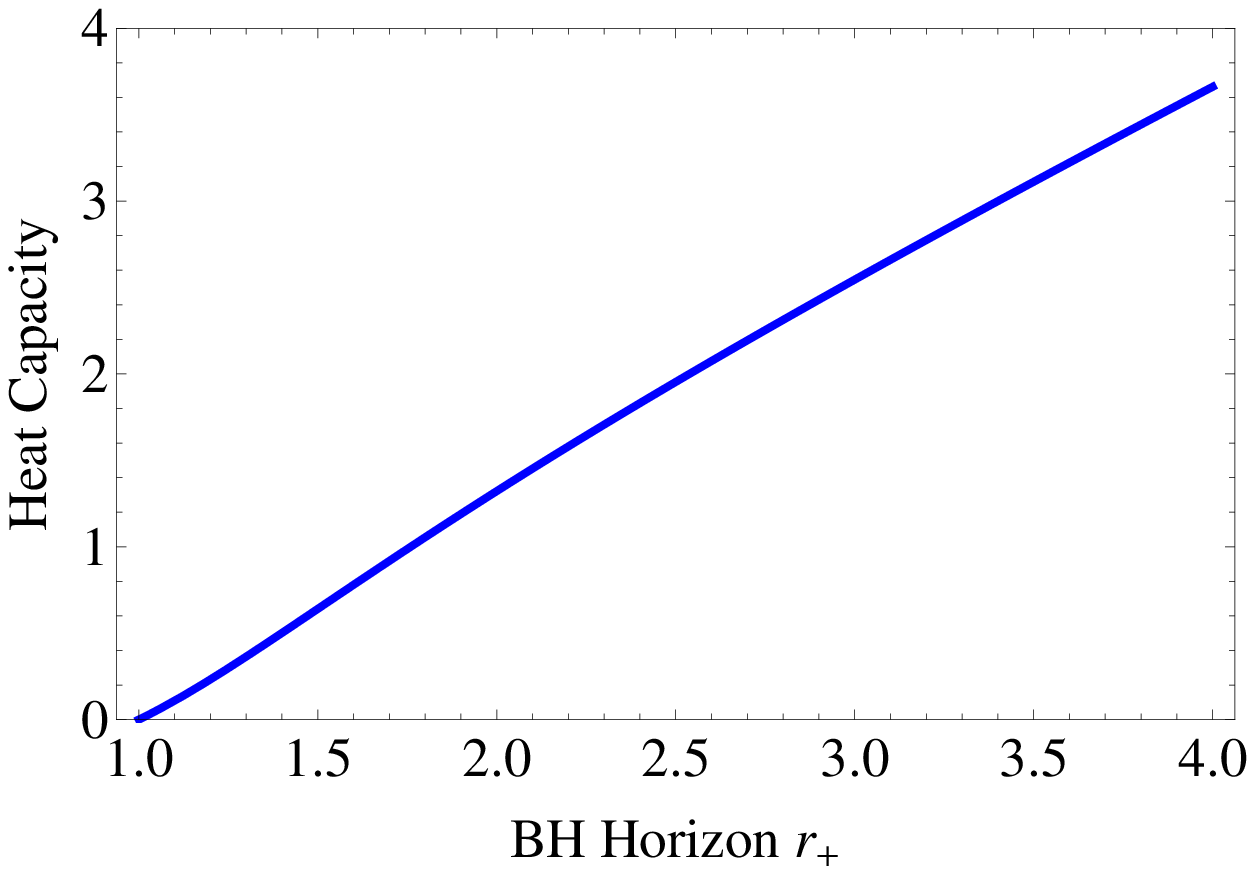}
      \caption{\label{FigTermodinamica} \textbf{Top panel:} The Mass Parameter. \textbf{Middle panel:} The Temperature. \textbf{Bottom Panel:} The Heat Capacity. All these plots were obtained by considering $l=L=\bar{G}=1$}. 
   \end{center}
\end{figure}

In figure \ref{FigTermodinamica} are shown the behavior of the mass parameter (top panel), temperature (middle panel) and heat capacity (bottom panel). It is direct to check that this behavior is generic for other values of the parameter $l$.

The equation \eqref{solucion} can take the form $f(r)=0$, which taking arbitrarily fixed $l$ and $L$, can have zero roots, two roots or one root, depending on the value of the mass parameter $M$.

In the top panel of figure \ref{FigTermodinamica}, following the analysis of references \cite{Dymnikova:2010zz,Aros:2019auf}, we plot the mass parameter $M(R)$, where $r=R$ corresponds to the root of the equation $f(r)=0$. $R$ can take the values $r_-$ and $r_+$, which correspond to the internal horizon and the black hole horizon, respectively. 
There is a critical value of the mass parameter $M_*=M(r_*)$, which corresponds to the minimum
value on the curve, and where there is an extremal black hole. At this point the internal horizon $r_-$ and the black hole horizon $r_+$ coincide, {\it i.e.} $r_*=r_-=r_+$. For $M<M_*$ the solution does not have horizons and for $M>M_*$ the solution has both internal and black hole horizon.

In the middle panel of figure \ref{FigTermodinamica}, we note that the temperature is vanishing just in the extremal black hole. This does not have a linear dependence on $r_+$ as the spinless BTZ solution \cite{Cruz:2004vp} and its dependence on $r_+$ also differs from the rotating BTZ solution \cite{Banados:1992wn}.

In the bottom panel of figure \ref{FigTermodinamica}, is displayed the heat capacity, which is always positive, therefore the solution is always stable. Its behavior also differs from the BTZ solution whose heat capacity is constant for the static case \cite{Cruz:2004vp}.

\section{Discussion and Conclusions}
We have obtained a new regular black hole solution in $(2+1)$ dimensions with presence of matter fields in the energy momentum tensor. 
For this, we have proposed a new model of energy density following the requirements described in reference \cite{Aros:2019quj}. This energy density has a finite value at the center, avoiding the formation of a central singularity. 

Our solution differs from BTZ-like geometry of reference \cite{Boehmer:2020hkn}, because in our solution is present  the mass function, unlike this reference, where is only present the constant mass parameter. Furthermore, in reference  \cite{Boehmer:2020hkn} the BTZ horizon is characterized for small values of the mass and cosmological constant. In our model the energy density tends to zero at infinity, equation \eqref{limiterho}, this differs from the model of reference \cite{Cruz:2004tz}, where the energy density diverges at infinity. Furthermore, our solution differs from the reference  \cite{Contreras:2019iwm}, which is based in the matching between the internal geometry obtained by means the Gravitational Decoupling algorithm and the external BTZ geometry.

The inclusion of our type of energy density plus a negative cosmological constant allows that the solution can have both flat as de--Sitter or Anti de--Sitter core, depending on the values of the $l$ and $L$. This latter is a proper characteristic of our solution, because other models of regular black holes have only a single type of core.

Regular solutions with a flat core have been studied in $(3+1)$ dimensions with a non--linear electromagnetic source in references \cite{Culetu:2014lca,Balart:2014cga}. However, for the $(2+1)$ case, with a nonzero energy density, the flat core is a proper feature of our solution. In $(3+1)$ dimensions, regular black hole solutions with a dS core have been widely studied in the literature. Our model with a dS core could be viewed as a $(2+1)$ extension of the Hayward metric. Since the Hayward model has been used to testing quantum gravity effects \cite{Rovelli:2017zoa}, our model could serve to study these effects in elsewhere due to the simplicity of the equations of motion in $(2+1)$ dimension. Furthermore our solution admits a core with AdS structure, which is a not studied feature so far. 

Using the recently definition of conserved charges of references \cite{Aoki:2020prb,Sorge:2020pdj} we have computed the total energy of our regular black hole solution, which is equal to the mass parameter $M$. The effectiveness of the definition used for computing the mass of our $(2+1)$ regular black hole model suggests that this formalism could be used to compute the mass of another models of regular black holes in $d \ge 4$ dimensions with presence of matter in the energy momentum tensor, as for example the models of references  \cite{Hayward:2005gi,DeLorenzo:2014pta,Rovelli:2014cta,Dymnikova:1992ux,Aros:2019quj}. 

The structure of the first law of thermodynamic for regular black holes is modified by the presence of matter fields in the energy momentum tensor \cite{Ma:2014qma}. So, we have provided a new version of the first law of thermodynamic \eqref{duTdS}, where it is defined a local definition of the variation of the energy, namely $du$, and, where the values of entropy and temperature are consistent with the previously known \cite{Banados:1992wn}. 
 We have showed that the sign of the local variation of the energy $du$ always coincides with the sign of variation of the total energy $dM$. Furthermore, at the hypothetical limit when the horizon radius $r_+ \to \infty$ the usual first law $dM=TdS$ is recovered.

\section*{Acknowledgements}
F. Tello-Ortiz acknowledges financial support by
CONICYT PFCHA/DOCTORADO-NACIONAL/2019-21190856  project ANT–1756  at
Universidad de Antofagasta, Chile. F. Tello-Ortiz thanks
the PhD program Doctorado en F\'{i}sica menci\'{o}n en F\'{i}sica
Matem\'{a}tica de la Universidad de Antofagasta for continuous support and encouragement.

\bibliography{epl}

\end{document}